%% file: main.tex
\documentclass[journal]{IEEEtran}
\usepackage{multicol}
\usepackage{lipsum,graphicx}
\usepackage{float}
\usepackage{epstopdf}
\usepackage{ifthen}
\usepackage[latin1]{inputenc}

\usepackage{amssymb}
\usepackage[cmex10]{amsmath}
\usepackage{dcolumn}
\usepackage{fixltx2e}
\usepackage[nolist]{acronym}

\DeclareGraphicsExtensions{.pdf,.jpeg,.png,.eps}
\usepackage{booktabs}

\usepackage[caption=false,font=footnotesize]{subfig}
\usepackage{comment}
\usepackage{color}

\usepackage{setspace}
\usepackage{mathrsfs}          
\usepackage{amsmath, amssymb, amsthm}
\usepackage{amsfonts}
\usepackage{mathrsfs}
\usepackage{multirow}        
\usepackage{framed}
\usepackage{psfrag}
\usepackage{graphicx}
\usepackage{caption}
\usepackage{units}            
\usepackage{algpseudocode} 

\input{mycommands}

\usepackage{lipsum}
\usepackage{mathtools}
\usepackage{cuted}

\begin{document}
	\input{abbreviations.tex}
	\title{Unified Low Complexity Radix-2 Architectures for  Time and Frequency-domain GFDM Modem
		\thanks{ This work has received funding from the European Union's Horizon 2020 research and innovation program under grant agreements No 777137 (5GRANGE project) and  No 732174 (ORCA Project [https://www.orca-project.eu/]).}}
	
	\author{
		\IEEEauthorblockN{Ahmad Nimr\IEEEauthorrefmark{1}, Marwa Chafii\IEEEauthorrefmark{2},  Gerhard Fettweis\IEEEauthorrefmark{1}}
		
		\IEEEauthorblockA{\IEEEauthorrefmark{1}Vodafone Chair Mobile Communication Systems, Technische Universit\"{a}t Dresden, Germany}
		
		\IEEEauthorblockA{\IEEEauthorrefmark{2} ENSEA, ETIS, Université Paris-Seine, CNRS, France}
		
		\IEEEauthorblockA{\small\texttt{ahmad.nimr@ifn.et.tu-dresden.de, marwa.chafii@ensea.fr, gerhard.fettweis@tu-dresden.de}}}	
	\maketitle
	\IEEEpeerreviewmaketitle
	
\input{0_abstract}
\input{1_introduction}
\input{2_section1}
\input{2_section2}

\input{3_section2}

\input{4_section3}
\input{5_section4}
\input{6_conclusions}
\bibliographystyle{IEEEtran}
\bibliography{references}
\end{document}

%% file: mycommands.tex
\usepackage{bm}

\newcommand{\LeftP} {\left\lbrace }
\newcommand{\RightP}{\right\rbrace }
\newcommand{\LeftPD} {\left( }
\newcommand{\RightPD}{\right)}

\newcommand{\compl}{\mathbb{C}}         
\newcommand{\intg}{\mathbb{N}}          

\newcommand{\ma}  [1]{ \bm{#1} } 
\newcommand{\mav}  [1]{ \bm{#1} } 
\newcommand{\IndexM}[3]{\left[ #1\right]_{\LeftPD #2,#3 \RightPD } } 
\newcommand{\IndexV}[2]{\left[ #1\right]_{\LeftPD #2 \RightPD } } 

\newcommand{\dft} [1] {\tilde{#1} } 

\newcommand{\Vect}  [1] {\mathrm{vec}  \LeftP #1 \RightP } 

\newcommand{\unvec} [3] {\mathrm{unvec}_{#2 \times #3} \LeftP #1 \RightP }

\newcommand{\modulo }[2] {<#1>_{#2} }

\newcommand{\DFT} [1] {\ma{F}_{#1}} 

\newcommand{\set} [1]{{\mathcal {#1}}} 
\newcommand{\Kon} {\set{K}_{\text{on}}} 
\newcommand{\Mon} {\set{M}_{\text{on}}} 
\newcommand{\Non} {\set{N}_{\text{on}}} 

\newcommand{\Tsub} {T_{\text{sub}}} 
\newcommand{\Tcp} {T_{\text{cp}}} 
\newcommand{\Tcs} {T_{\text{cs}}} 
\newcommand{\Tt} {T_{\text{t}}} 

\newcommand{\V}  [3]{ \ma{V}^{(#3)}_{#1,#2} } 
\newcommand{\Z}  [3]{ \ma{Z}^{(#3)}_{#1,#2} } 
\newcommand{\Zbar}  [3]{ \bar{\ma{Z}}^{(#3)}_{#1,#2} } 



%% file: abbreviations.tex
\begin{acronym}
	\acro{FD}{frequency domain}
	\acro{TD}{time domain}
	\acro{OOB}{out-of-band}
	\acro{RRC}{root-raised Cosine}
	\acro{RC}{raised-cosine}
	\acro{ISI}{inter-symbol-interference}
	\acro{ZF}{zero-forcing}
	\acro{MF}{matched filter}
	\acro{SINR}{signal-to-interference-plus-noise ratio}
	\acro{SNR}{signal-to-noise ratio}
	\acro{FIR}{finite impulse repose }
	\acro{DFT}{discrete Fourier transform}
	\acro{OFDM}{orthogonal frequency division multiplexing}
	\acro{GFDM}{generalized frequency division multiplexing}
	\acro{ICI}{inter-carrier-interference}
	\acro{IAI}{inter-antenna-interference}
	\acro{NEF}{noise-enhancement factor}
	\acro{FDE}{frequency fomain equalization}
	\acro{SVD}{singular-value decomposition}
	\acro{AWGN}{additive white Gaussian noise}
	\acro{DTFT}{discrete-time Fourier transform}
	\acro{FFT}{fast Fourier transform}
	\acro{SIR}{signal-to-interference ratio}
	\acro{DZT}{discrete Zak transform}
	\acro{MIMO}{multiple-input multiple-output}
	\acro{PAPR}{peak-to-average power ratio}
	\acro{F-OFDM}{filtered OFDM}
	\acro{CP}{cyclic prefix}
	\acro{CS}{cyclic suffix}
	\acro{ZP}{zero padding}
	\acro{IBI}{inter-block-interference}
	\acro{GT}{guard tone}
	\acro{UF-OFDM}{universal-filtered OFDM}
	\acro{FBMC}{filter bank multicarrier}
	\acro{OQAM}{offset quadrature amplitude modulation}
	\acro{FER}{frame error rate}
	\acro{MMSE}{minimum mean square error}
	\acro{IAI}{inter-antenna-interference}
	\acro{MCS}{modulation coding scheme}
	\acro{PSD}{power spectral density}
	\acro{IoT}{Internet of Things}
	\acro{MTC}{machine-type communication}
	\acro{STC}{space-time coding}
	\acro{TR-STC}{time-reversal space-time coding}
	\acro{MRC}{maximum-ratio combiner}
	\acro{LS}{least squares}
	\acro{LMMSE}{linear minimum mean squared error}
	\acro{CIR}{channel impulse response}
	\acro{STO}{symbol time offset}
	\acro{CFO}{carrier frequency offset}
	\acro{UE}{user equipment}
	\acro{FO}{frequency offset}
	\acro{TO}{time offset}
	\acro{BS}{base station}
	\acro{FMT}{filtered multitone }
	\acro{DAC}{digital-to-analogue converter }
	\acro{FO}{frequency offset}
	\acro{TO}{time offset}
	\acro{ISI}{inter-symbol-interference}
	\acro{IUI}{inter-user-interference}
	\acro{IBI}{inter-block-interference}
	\acro{i.i.d.}{independent and identically distributed}
	\acro{SER}{symbol error rate}
	\acro{LTE}{Long Term Evolution}
	\acro{SISO}{single-input single-output}
	\acro{Rx}{receive}
	\acro{Tx}{transmit}
	\acro{MSE}{mean squared error}
	\acro{IFPI}{interference-free pilot insertion}
	\acro{PDP}{power-delay-profile}
	\acro{ML}{maximum likelihood}
	\acro{5G}{5th generation}
	\acro{4G}{4th generation}
	\acro{NR}{New Radio}
	\acro{eMBB}{enhanced media broadband}
	\acro{URLLC}{ultra-reliable and low-latency communication}
	\acro{mMTC}{massive machine type communication}
	\acro{SDR}{software defined radio}
	\acro{RF}{radio frequency}
	\acro{PHY}{physical layer}
	\acro{MAC}{medium access layer}
	\acro{FPGA}{field programmable gate array}
	\acro{IDFT}{inverse discrete Fourier transform}
	\acro{DRAM}{dynamic random access memory}
	\acro{BRAM}{block RAM}
	\acro{FIFO}{first in first out}
	\acro{D/A}{digital to analog}
\acro{EVA} {extended vehicular A channel model} 
	\acro{OTFS}{Orthogonal time frequency space modulation}
	\acro{SFFT}{symplectic finite Fourier transform}

	\acro{ACLR}{adjacent channel leakage rejection}
	\acro{ADC}{analog-to-digital converter}
	\acro{AGC}{automatic gain control}
	\acro{CEP}{channel estimation preamble}
	\acro{DPD}{digital pre-distortion}
	\acro{PA}{power amplifier}
		\acro{LTV}{linear time-variant}
	
		\acro{NMSE}{normalized mean-squared error}
	\acro{PRB}{physical resource block}
	\acro{BER}{bit error rate}
		\acro{FER}{frame error rate}
	\acro{DL}{downlink}
	\acro{UL}{uplink}
	\acro{FO}{frequency offset}
	\acro{TO}{time offset}
	\acro{MA}{multiple access}

	\acro{INI}{inter-numerology-interference}
	\acro{PCCC}{parallel concatenated convolutional code}
	\acro{CCDF}{complementary cumulative distribution function}
	\acro{SC}{single carrier}
	\acro{FDMA}{frequency division multiple access}
	\acro{IP}{intellectual property}
	\acro{CM}{complex multiplication}
	\acro{DSP}{digital signal processor }
\acro{LUT}	{lookup table}
\acro{RAM}	{random-access memory}
\acro{RW}	{read-and-write}
\acro{R/W}	{read-or-write}
\acro{MCM}{multicarrier modulation}
\end{acronym}

%% file: 0_abstract.tex
\begin{abstract}
Most of the conventional multicarrier waveforms explicitly or implicitly involve a \ac{GFDM}-based modem as a core part of the baseband processing. Some are based on \ac{GFDM} with a single prototype filter, e.g. \ac{OFDM} and others employ multiple filters such as \ac{FBMC}. Moreover, the  \ac{GFDM} degrees of freedom combined with multiple prototype filters design allow the development and optimization of new waveforms. Nevertheless, \ac{GFDM} has been widely considered as a complex modulation because of the requirements of odd number of  subcarriers or subsymbols. Accordingly, the current state of the art implementations consume high resources. One solution to reduce the complexity  is  utilizing radix-2 parameters.  Due to the advancement in \ac{GFDM} filter design, the constraint of using odd parameters has been  overcome and radix-2 realization is now possible.
In this paper, we propose a unified low complexity  architecture that can be reconfigured to provide both time-domain and frequency-domain modulation/demodulation. The design consists of several radix-2 \ac{FFT} and memory blocks, in addition to  one complex multiplier. Moreover, we provide a unified architecture for the state of the art implementations,  which is designed based on direct computation of circular convolution using  parallel multiplier chains. As we demonstrate in this work, the \ac{FFT}-based architecture is computationally more efficient, provides more flexibility, significantly reduces the resource consumption, and achieves similar latency for larger block size. 
\end{abstract}

%% file: 1_introduction.tex
\acresetall
\section{Introduction}\label{sec:introduction}
Multicarrier modulation is proposed as an alternative to single-carrier modulation to enable  low complexity receiver design, especially in frequency selective channels \cite{MCM}. \ac{OFDM} is  the dominant scheme adopted by various standards for wired and wireless systems, such as television and audio broadcasting, DSL, wireless area networks, and 4G mobile communications \cite{OFDM}. The main advantage of \ac{OFDM} is the low complexity implementation. However, \ac{OFDM} has well-known problems that limit its usage in various applications \cite{5GNow}. For instance,  because of its sensitivity to frequency misalignment, \ac{OFDM} is not suitable for massive networks, that require asynchronous multiple access in order to get rid of synchronization overhead. The high \ac{OOB} emissions of \ac{OFDM} reduce its efficiency  for dynamic spectrum access. Moreover, the high \ac{PAPR} complicates the radio frequency design and increases the cost  when deployed in machine type communications. To tackle these challenges, additional processing techniques on top of \ac{OFDM} have been suggested. In \cite{Continuous_OFDM}, the \ac{OFDM} signal is smoothed by means of precoding to reduce the \ac{OOB}. However, this solution further complicates the system implementation. Other low complex variants such as windowed-\ac{OFDM} \cite{medjahdi2017wola}, and  filtered-\ac{OFDM} \cite{wild20145g} enhance the overall \ac{OFDM} signal by introducing additional processing after the \ac{OFDM} modulator. In addition to \ac{OFDM} and its variants, new modulation techniques are proposed to attain the requirements of new use cases.  Some aim at providing very low \ac{OOB} emissions and resistance to frequency and time misalignments, e.g. \ac{FBMC} \cite{FBMC}, while  
others target the PAPR reduction, e.g. DFT-spread-OFDM
\cite{DFT-S-OFDM}.  Furthermore, \ac{GFDM} as a waveform,  is  proposed in \cite{GFDM} as an alternative to \ac{OFDM} to improve the spectral efficiency by reducing the \ac{CP} overhead of \ac{OFDM}. One \ac{GFDM} block can have a duration of several \ac{OFDM} symbols with one \ac{CP} rather than employing a \ac{CP} per each \ac{OFDM} symbol. On the other hand, \ac{GFDM} provides more degrees of freedom that allow the design of the waveform depending on the use case. For example,  targeting high throughput, the design in \cite{GFDM_Design} considers a periodic  Raised-Cosine with high roll-off factor, which is well-localized in the \ac{TD}. For smooth transitions between the blocks, the first and last subsymbols are not used for data transmission but  rather for carrying pilot symbols. To preserve a cyclic structure that enables \ac{FD} equalization, a unique word prefix is inserted at the beginning of the frame. With  this configuration, \ac{GFDM} achieves very low \ac{OOB}, which reduces the number of guard subcarriers for a given spectrum mask compared to \ac{OFDM}. As a result, a significant throughput gain is  archived. Nevertheless, all these benefits are at the cost of increasing the  complexity of the receiver because  of sacrificing the orthogonality.

The common thread among all multicarrier techniques is the transmission of data in parallel streams, which are then  superimposed to formulate the final signal. Each stream can be seen as a single-carrier modulation with a specific pulse shape. In the state of the art multicarrier systems, the pulse shapes are generated by a shift of the prototype filter\footnote{Prototype pulse and prototype filter are used exchangeably  in this paper.} in the \ac{FD}  and each stream is denoted as subcarrier. These systems correspond to one-dimensional filter design.  For instance, \ac{OFDM} uses a rectangular pulse shape of length equal to the  \ac{OFDM} symbol duration, while \ac{FMT} employs an optimized pulse shape  that is several times longer, where multiple symbols overlap \cite{FMT}. 
In addition to an \ac{FD} shift, a \ac{TD} shift of the prototype pulse is possible, and refers to the subsymbol concept as introduced by \ac{GFDM}. Actually, even the multicarrier based on the frequency shift can be reformulated in a \ac{GFDM} form \cite{gaspar2015gfdm}.
 On the other hand,  some modulation schemes, such as \ac{FBMC} \cite{OFDMvsFBMC}, implicitly employ two prototype pulses. These schemes belong to two-dimensional filter design. The general case is a special form of the multidimensional wave principle introduced by Alfred Fettweis in \cite{multi_dimensional_waveform}, where the waveform is generated from multiple prototype pulses. As shown in this paper,  \ac{GFDM} appears as a basic building unit of all previously cited multicarrier waveforms and can be used to invent optimized waveforms that employ multiple prototype pulses. Therefore, a low-complex implementation is essential for the development of flexible multicarrier systems.

As we show in this paper, the \ac{GFDM} modem can be implemented with \ac{DFT} blocks. If the involved \ac{DFT} sizes are of radix-2 basis, i.e. power of $2$, then \ac{DFT} can be computed with a \ac{FFT} algorithm. For example in the famous Cooley-Tukey implementation \cite{cooley1965algorithm},  the input vector of size $N$ is split into two sub-vectors of size $N/2$ with respect to the even and odd indexes. The $N/2$-\ac{DFT} of each sub-vector is computed using \ac{FFT}. The $N$-\ac{DFT} is achieved  by combining the \acp{DFT} of the sub vectors according to the butterfly diagram with the corresponding twiddle factors. This combination requires $N$ complex multiplications. The recursion is repeated until reaching the size $2$, where $2$-\ac{DFT} is computed with an addition and a subtraction. Therefore, radix-2 \ac{FFT} significantly reduces the complexity from $\set{O}{(N^2)}$ to $\set{O}{(N\log_2N)}$ complex multiplications.  

As a consequence of the assumption of real-valued symmetric prototype pulse, it is conventionally considered that the number of subcarriers and subsymbols in \ac{GFDM} should not be both even number to not degrade the overall performance \cite{Gabor}. This assumption leads to a complicated implementation architecture that considers only one radix-2 parameter. The \ac{TD} real-time implementation in \cite{danneberg2015flexible} is based on radix-2 number only for subcarriers, while the \ac{FD} \cite{FD-GFDM} considers only radix-2 number of subsymbols. \ac{CP} is usually added to the \ac{GFDM} block to enable \ac{FD} channel equalization, which requires an additional \ac{DFT} transform. When \ac{TD} modulator  and \ac{FD} demodulator are  used, only one additional \ac{DFT} is needed for the channel equalization. Nevertheless, with the previous assumption, either \ac{TD} or \ac{FD} is allowed at the same time.  Due to the advancement of filter design \cite{nimr2017optimal},  the condition of using only odd parameters is no longer necessary. Therefore, rethinking  efficient implementation is required.  \\
\indent The proposed low-complex architecture in this paper provides a unified architecture for \ac{TD} and \ac{FD} realizations of \ac{GFDM} modulation and demodulation. This architecture employs several \ac{FFT} blocks, several memory blocks and only one multiplier, which significantly reduces the required resources for hardware implementation. Additionally, the flexible proposed design enable further extension of \ac{GFDM} to generate coded \ac{OFDM} waveforms such as \ac{OTFS}\cite{OTFS_wcnchadani2017orthogonal}. In addition, we review the state of the art \ac{GFDM} implementations and  show that the different \ac{TD} and \ac{FD} implementations  can actually be realized with another  unified  architecture. Namely, this design requires two \ac{FFT} blocks, several parallel chains of complex multipliers and memory blocks.

The remainder of the paper is organized as follows: Section \ref{sec: multicarrier framework}  provides a general representation of multicarrier systems highlighting the importance of \ac{GFDM} as a building block. Section \ref{sec:GFDM represnetation} proposes  an advanced representation of \ac{GFDM} in order  to give a closer insight of its structure. The proposed hardware architecture is introduced in \ref{sec:prposed archetecture}. In Section \ref{sec:GFDM modem implementation}, we reproduce the state of the art implementations in a unified architecture. Section \ref{sec:complexity anlysis} is dedicated for the complexity analysis  with respect to software and hardware implementation. Finally, Section \ref{sec:conclusions} concludes the paper.

The following notations are used throughout the paper: scalars are represented with italic letters $a$, $A$.  Column vectors, matrices are denoted  with bold-face letters in lower-case for vectors $\ma{a}$, upper-case for matrices $\ma{A}$. The field of complex numbers is denoted as $\compl$ and the finite set as calligraphic face $\set{A}$, with  $|\set{A}|$ is the number of elements. The  $(m,n)$-th element of a matrix is given by $\IndexM{\ma{A}}{m}{n}$ and the $n$-th column of $\ma{A}$ by $\IndexM{\ma{A}}{:}{n}$. 
We use $\{\cdot\}^T$, $\{\cdot\}^H$ for matrix transpose, and  Hermitian transpose. Moreover, $\ma{a} = \Vect{\ma{A}}$ indicates vectorization of $\ma{A} \in \compl^{K\times M}$, whereas its inverse operator is expressed by $\unvec{\ma{a}}{K}{M}$. The symbols  $\otimes$ and $\odot$ denote Kronecker and element-wise products. Finally, the modulo-$N$ operation is represented as  $\modulo {\cdot}{N}$.

%% file: 2_section1.tex
\section{Multicarrier waveforms overview}\label{sec: multicarrier framework}
In  linear \ac{MCM}, a stream of data symbols is split into $N$ parallel substreams. Let $d_{k,m}$ be the stream corresponding to the $k$-th  \emph{subcarrier}  and the $m$-th \emph{subsymbol}. Assuming  $K$ subcarriers and $M$ subsymbols,  {$N=KM$}. Each substream is  modulated with a transmitter pulse $g^{(\text{tx})}_{k,m}(t)$. In the conventional \ac{MCM} techniques, 
the pulses $\{g^{(\text{tx})}_{k,m}\}$ have finite length $T_t$, which enables the definition of finite-length modulated symbol as
\begin{equation}
\small
\begin{split}
x_{i}^{(\text{tx})}(t) = \sum_{k\in \Kon}\sum_{m\in \Mon}d_{k,m,i}g^{(\text{tx})}_{k,m}(t),~ t\in [0, T_t],
\end{split}
\end{equation}
where $\Kon$ and $\Mon$ denote the sets of active subcarriers and subsymbols. The transmitted signal is generated by \emph{multiplexing} the individual blocks with spacing interval $T_s$, so that $1/T_{s}$ is the data rate per stream,
\begin{equation}
\small
\begin{split}
x^{(\text{tx})}(t) &= \sum_{i\in \intg}x_{i}^{(\text{tx})}(t-iT_s).
\end{split}
\end{equation}
The time difference $T_o = T_t-T_s$ defines an overlapping ($T_o>0$) or a guard ($T_o\leq 0$) interval between successive blocks. For instance, in typical \ac{FBMC} \cite{fbmc-primer}, the overlapping factor is $4$ which means that $T_t = 4T_s$, and thus $T_o = 3T_s$.
In some multicarrier waveforms like \ac{OFDM} and \ac{GFDM}, the transmitted block is generated form additional processing on top of a \emph{core block} $x_i(t)$ of duration $T$. These include adding \ac{CP} and \ac{CS} overheads of duration $\Tcp $ and $\Tcs$  respectively, in order to facilitate simple \ac{FD} channel equalization and compensate for the time offset. Afterwards,  windowing or filtering can be applied to reduce \ac{OOB}. In the case of windowing, a window $w(t)$ of duration $\Tt = T+ \Tcp +\Tcs$ is applied such that
\begin{equation}
\small
\begin{split}
x_{i}^{(\text{tx})}(t) &= w(t)x_i(\modulo {t-\Tcp}{T}).
\end{split}
\end{equation}
With filtering the block duration is extended by the filter tail $T_f$, i.e. $\Tt = T+ \Tcp +\Tcs + T_f $. Let $w(t)$ be the filter impulse response, the transmitted block is expressed as
\begin{equation}
\small
\begin{split}
x_{i}^{(\text{tx})}(t) &= w(t)*x_i(\modulo {t-\Tcp}{T}).
\end{split}
\end{equation}
Accordingly, we define core pulses $\{g_{k,m}(t)\}$ of duration $T$, which are used to generate the modulation pulses $\{g^{(\text{tx})}_{k,m}\}$ depending on the windowing or filtering, such that
 \begin{equation}
\small
g^{(\text{tx})}_{k,m}(t)  = w(t)[\cdot~\mbox{or}~ *] g_{k,m} (\modulo {t-\Tcp}{T}).
\end{equation}
Therefore, the core block can be expressed as 
\begin{equation}
\small
x_i(t) = \sum\limits_{k\in\Kon}\sum\limits_{m\in\Mon}d_{k,m,i}g_{k,m}(t), t\in [0, T].
\end{equation}
\subsection{Discrete-time representation}
Without loss of generality, the discrete-time signal is obtained by the sampling of the continuous-time signal with frequency $F_s = \tfrac{N}{T}$. The samples of the core block $x_i[n] = x_i(\tfrac{n}{F_s})$ are given by
\begin{equation}
\small
x_i[n] = \sum_{m\in \Kon}\sum_{m\in \Mon}d_{k,m,i}g_{k,m}[n], n = 0\cdots N-1.
\end{equation}
This linear relation  can be reformulated in a matrix form $\ma{x}_i = \ma{A}\ma{d}_i,$  where 
{$\IndexV{\ma{d}_i}{k+mK} = d_{k,m,i}, ~(k,m)\in 
	\Kon \times \Mon$ and $0$ elsewhere.} $\ma{A}\in \compl^{N\times N}$ is the modulation matrix defined by $\IndexM{\ma{A}}{n}{k+mK} = g_{k,m}[n]$.
\begin{figure*}[t]	
	\centering
	\includegraphics[width=1\textwidth]{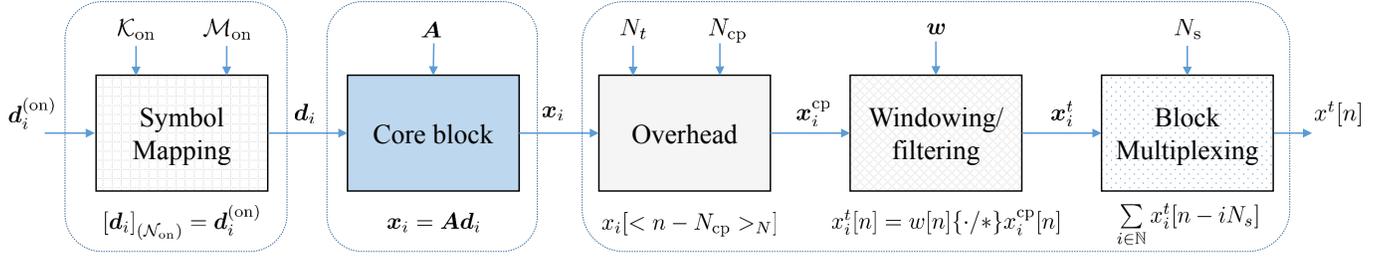}
	\caption{Multicarrier waveforms generator stages.}\label{fig:multicarrier_tx}
\end{figure*}
Based on that, the multicarrier waveform realization can be split into three independent modules as depicted in Fig.~\ref{fig:multicarrier_tx};
\begin{enumerate}
\item Symbol mapping: a vector of data symbols $\ma{d}_i^{(\text{on})} \in \compl^{|\Non|\times 1}$ is mapped to $\ma{d}_i$, such that $\IndexV{\ma{d}_i}{\Non} =  \ma{d}_i^{(\text{on})}$, 
where $
\small
\Non = \left\lbrace n = k+mK,~ (k,m) \in \Kon\times \Mon  \right\rbrace \label{eq: active n}.
$
\item Core block modulation.
 \item Further processing and multiplexing.
\end{enumerate}
The core block implementation is the essential part. In the most general cases it requires $N^2$ complex multiplications and a memory to store $N^2$ complex coefficients. In practical \ac{MCM}, the matrix $\ma{A}$ has a well-defined structure based on the design of the pulses $\{{g_{k,m}[n]}\}$ which define the columns of $\ma{A}$. This structure can be exploited in the implementation.
\subsection{Modulation pulses design}\label{sec:pulse shape design}
\begin{figure}[t]	
	\centering
	\includegraphics[width=.5\textwidth]{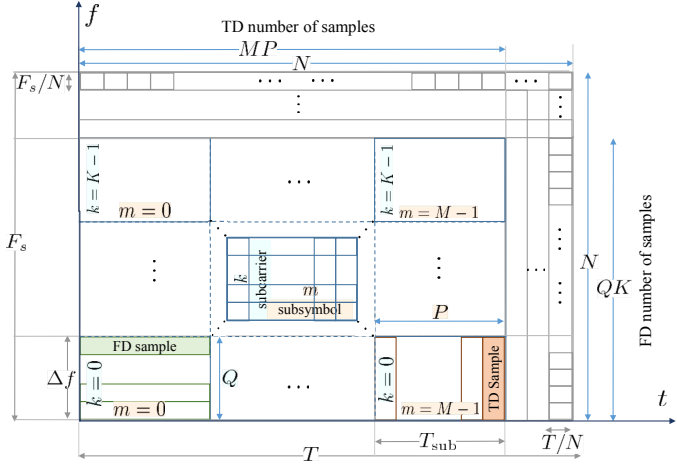}
	\caption{Time-frequency design parameters.}\label{fig:Sampling}
\end{figure}

\begin{figure}[h]	
	\centering
	\includegraphics[width=.5\textwidth]{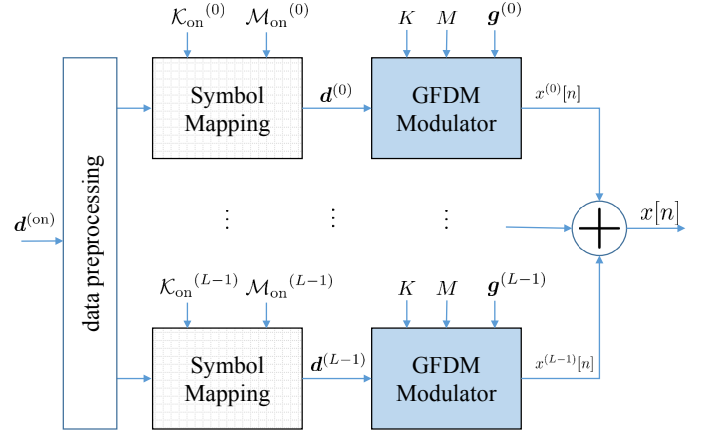}
	\caption{Core block with multiple prototype pulses.}\label{fig:multi_pulse}
\end{figure}

It is common to derive $\{g_{k,m}(t)\}$ from one prototype pulse  $g(t)$ by means of shift in the time and frequency domains.
Let $\Delta f$, $\Tsub$ be the \ac{FD}  subcarrier spacing and \ac{TD} subsymbol spacing, respectively. Then,
\begin{equation}
\small
g_{k,m}(t) = u_T(t)g(t-m\Tsub) e^{j2\pi \Delta f k t }.
\end{equation}
where $u_T(t)$ is a rectangular window of duration $T$  used to confine the pulse shape in the time duration $T$. In order to preserve the energy per stream, the pulse $g(t)$ needs to be periodic. An explicit periodic prototype pulse is used, e.g. in  \ac{GFDM} and \ac{OFDM}, whereas \ac{FMT} \cite{FMT}, which originally defines only subcarriers, can be reformulated to involve subsymbols. The set $\Mon$ is determined based on the filter overlapping factor. Under this constraint, we focus on a periodic prototype pulse.
For practical implementation, we assume that  $Q = {T\Delta f}$ and $P = {\Tsub}F_s = \tfrac{N}{T}{\Tsub} $  are integer numbers so that
\begin{equation}
\small
g_{k,m}[n] = g[\modulo {n-mP}{N}] e^{j2\pi \frac{nkQ}{N} }. \label{eq:general form of GFDM}
\end{equation}
The parameters $Q$ and $P$ denote the subcarrier and subsymbol spacing in samples, respectively (see Fig.~\ref{fig:Sampling}).
Actually $\{g_{k,m}\}$ belong to Gabor time-frequency lattices \cite{daubechies1994gabor}, with  $g_{k\alpha,m\beta}[n] = g[\modulo {n-m\beta}{N}] e^{j2\pi nk\alpha }$, $\alpha = Q/N,~ \beta = P$. In order to uniquely demodulate the data symbols $d_{k,m}$ from a given $x[n]$, Wexler-Raz duality condition  \cite{WEXLER1990207} must be satisfied. Thus, there exists a pulse $h[n]$ that attains  $\langle h, g_{k/\beta,m/\alpha}\rangle = \alpha \beta \delta_{k,0}\delta_{m,0}$, where $g_{k/\beta,m/\alpha}$ is the dual Gabor lattice, $\delta_{ij}$ refers to  Kronecker delta,  and $\langle\cdot\rangle$ denotes the inner product. Furthermore, $h$ is the  demodulator prototype pulse,  such that  $d_{k,m} = \langle x, h_{k\alpha, m\beta}\rangle$. It  is rigorously  proven in \cite{janssen1994signal} that Wexler-Raz duality condition cannot be fulfilled if $\alpha\beta >1$.
Based on that the choice of $P$ and $Q$ is influenced by
\begin{itemize}
	\item  $QP\leq N$, which is necessary but not sufficient to achieve Wexler-Raz duality condition.
	\item $\Delta f K \leq F_s = \tfrac{N}{T}$, which is a necessary, but not sufficient  condition for the signal to have a bandwidth $B\leq F_s$.
\end{itemize}
Consequently, the design needs to fulfill the conditions
\begin{equation*}
\small
Q\leq M,~ P \leq K.
\end{equation*} 
It is required for efficient implementation purposes to consider the  case $PQ = N$, i.e.   $P = K$ and $Q = M$, which correspond to critically sampled system. Essentially, this special case corresponds to \ac{GFDM}-based system \cite{GFDM}. Fortunately, a system that requires $PQ < N$, i.e. an over-sampled system,  can be redesigned to satisfy the condition $PQ = N$ by managing the sampling frequency, the block length, and properly defining the active sets. Accordingly, let $Q = T\Delta f  = M$ and $P = \Tsub F_s = \frac{K}{L}$, where $L$ is a positive integer. The set of active subcarriers $\Kon$ is adjusted to the available bandwidth $B\leq F_s$. The modulation pulses can be redefined as
\begin{equation*}
\small
\begin{split}
g_{k,mL + l}[n]
&= g(\modulo {n- mK -l\tfrac{K}{L}}{N}) e^{j2\pi \frac{nkM}{N}}.
\end{split}
\end{equation*}
Then, $L$  prototype pulses can be defined by
\begin{equation*}
\small
g^{(l)}[n] = g[\modulo {n- l\tfrac{K}{L}}{N}] , ~ l =0\cdots, L-1.
\end{equation*}
Each pulse shape is used to generate a subset of the pulse shapes with the subsymbol spacing $K$ and the set 
\begin{equation*}
\small
{\Mon}^{(l)} =\{mL + l, ~ l\leq mL+l <M\}.
\end{equation*}
The final block can be expressed as superposition of $L$ \ac{GFDM}-based block $x[n]= \sum_{l=0}^{L-1}x^{(l)}[n]$,
where 
\begin{equation}
\small
x^{(l)}[n] = \sum_{k\in \Kon}\sum_{m\in {\Mon}^{(l)}} d_{k,m}g^{(l)}[\modulo {n-mK}{N}] e^{j2\pi \frac{nk}{K} }.
\end{equation}
An alternative reformulation can be achieved in the frequency domain with respect to the $N$-DFT of ${g}_{k,m}[q]$ given by
\begin{equation}
\tilde{g}_{k,m}[q] = \tilde{g}[\modulo {q-kQ}{N}]e^{-j2\pi \frac{qmP}{N} }.\label{eq:general frequency domain }
\end{equation}
The design is adjusted such that $P = \Tsub F_s = K$ and $Q = \tfrac{M}{L}$ resulting in the prototype pulses 
\begin{equation*}
\tilde{g}^{(l)}[q] = \tilde{g}[\modulo {q- l\tfrac{M}{L}}{N}] , ~ l =0\cdots, L-1,
\end{equation*}
for the sets ${\Kon}^{(l)} =\{kL + l, ~ l\leq kL+l <K\}$. 
This approach can be generalized to design a  waveform with $L$ prototype pulses $\{g^{(l)}[n]\}$. Each prototype pulse is associated with the sets ${\Kon}^{(l)}$ and ${\Mon}^{(l)}$, as shown in Fig.~\ref{fig:multi_pulse}. Thus, for $(k,m) \in {\Kon}^{(l)}\times {\Mon}^{(l)}$
\begin{equation}
\small
g_{k,m}[n] =  g^{(l)}(\modulo {n- mK}{N}) e^{j2\pi \frac{nk}{K}}.
\end{equation}
The input data symbol can be preprocessed prior to modulation,  for instance to produce \ac{OQAM} \cite{OQAM}. This design allows the realization of wide range of multicarrier waveforms. For example, to generate \ac{FBMC}, where $Q = M$ and $P = K/2$, first, the complex QAM data symbols $d_{k,m} = d^{(I)}_{k,m}+ jd^{(Q)}_{k,m}$ are split into two \ac{OQAM} precoded streams, $d^{(0)}_{k,m} = \theta_{0,k} d^{(I)}_{k,m}$, $d^{(1)}_{k,m} = \theta_{1,k} d^{(Q)}_{k,m}$, where 
\begin{equation*}
\small
\theta_{0,k} = \left\lbrace\begin{array}{ll}
j,& ~ k \mbox{ is even}\\
1,& ~ k \mbox{ is odd}
\end{array}\right\rbrace , \theta_{1,k} = \left\lbrace\begin{array}{ll}
1,& ~ k \mbox{ is even}\\
j,& ~ k \mbox{ is odd}
\end{array}\right\rbrace .
\end{equation*}
The streams are then fed to two  \ac{GFDM} modulators with the parameters $g^{(0)}[n] = g[n]$,  $g^{(1)}[n] = g[<n-K/2>]$, ${\Kon}^{(0)} = {\Kon}^{(1)}$, and ${\Mon}^{(0)} = {\Mon}^{(1)}$. The output is a superposition of both \ac{GFDM} blocks, i.e. $x[n] = x^{(0)}[n]+x^{(1)}[n]$.
\subsection{Relation to multidimensional digital filtering}
\ac{GFDM} can be seen as a circular filtering of data symbols per subcarrier. For simplicity, consider a single carrier system, i.e., $K=1$. From Fig.~\ref{fig:multi_pulse},
\begin{equation}
x[n] = \sum_{l=0}^{L-1}\sum\limits_{m\in\Mon^{(l)}}d_m^{(l)}g^{(l)}[<n-m>_N].\label{eq:multipulse single carrier}
\end{equation}
The multidimensional circular filtering can be expressed as 
\begin{equation}
\small
\begin{split}
x[n_0,\cdots&, n_{L-1} ] =\sum_{m_0=0}^{M-1}\cdots \sum_{m_{L-1}=0}^{M-1}D_{m_0,\cdots, m_{L-1}}\\
&\times G[<n_0-m_0>_N, \cdots, <n_{L-1}-m_{L-1}>_N].
\end{split}
\end{equation}
Here, $D_{m_0,\cdots, m_{L-1}}$ corresponds to the multi-dimensional data, while $G[n_0, \cdots, n_{L-1}]$ is the multidimensional filter. Therefore, \ac{GFDM} with multiple prototype pulses is a special case, where  $D_{m_0,\cdots, m_{L-1}}$ is sparse with  $M$ non-zero values. The multidimensional filter coefficients are set to satisfy \eqref{eq:multipulse single carrier}. Although the work of Alfred Fettweis on multidimensional wave-digital principle \cite{multi_dimensional_waveform} targets other applications, it inspires more investigation for wireless communications. In this context, a one-dimensional circular filtering appears in the received signal model of \ac{OFDM} under frequency selective channel. On the other hand,  the received signal of \ac{OTFS} \cite{OTFS_wcnchadani2017orthogonal} is modeled as two-dimensional filtering with a time-variant channel response in the delay-Doppler domain. Accordingly, three-dimensional filtering can be a natural candidate when considering the spatial domain. As we show in this paper, a \ac{GFDM} core can be used to efficiently realize one-dimensional  convolution, which is a crucial step for multidimensional filtering.

%% file: 2_section2.tex
\section{GFDM time-frequency representations}\label{sec:GFDM represnetation}
\begin{table*}[h]
	\centering
	\small
	\caption{Summary of \ac{GFDM} equations.}\label{Tab: GFDM basics suammary}
	\begin{tabular}{l|c|c}	 		
		\small
		\rule{0pt}{10pt}& {Time domain}& {Frequency domain}\\
		\hline				
		  \multicolumn{3}{c}{	\rule{0pt}{10pt} Modulation}\\
		\hline
		\hline
		\rule{0pt}{15pt}Conv. &$\IndexM{\V{M}{K}{\ma{x}}}{p}{q}=
\sum\limits_{m=0}^{M-1}\IndexM{\V{M}{K}{\ma{g}}}{<p-m>_M}{q}\IndexM{\ma{D}^T\DFT{K}^H}{m}{q}$& $\IndexM{\V{K}{M}{\tilde{\ma{x}}}}{q}{p}
		= \sum\limits_{k=0}^{K-1}\IndexM{\V{K}{M}{\tilde{\ma{g}}}}{<q-k>_K}{p} \IndexM{\ma{D}\DFT{M}}{k}{p}$\\
		\hline 
		\rule{0pt}{14pt}Zak trans. &${\V{M}{K}{\ma{x}}} =
		\frac{1}{M K}\DFT{M}^H\left(\ma{W}_{\text{tx}}\odot\left[\DFT{K}^H{\ma{D}\DFT{M}}\right]\right)^T$& ${\V{K}{M}{\tilde{\ma{x}}}} =
		\frac{1}{K}\DFT{K}\left(\ma{W}_{\text{tx}}\odot\left[\DFT{K}^H{\ma{D}\DFT{M}}\right]\right)$\\
		\hline	
		\rule{0pt}{14pt}Tx win. & $\ma{W}_{\text{tx}} = K{\Z{M}{K}{\ma{g}}}^T$ & $\ma{W}_{\text{tx}} = K{\Zbar{K}{M}{\tilde{\ma{g}}}}$ \\
		\hline
			  \multicolumn{3}{c}{\rule{0pt}{10pt} Demodulation}\\		
		\hline 
		\hline 
		\rule{0pt}{14pt}Zak trans. &$\hat{\ma{D}} = \frac{1}{M }\DFT{K}\left(\ma{W}_{\text{rx}}\odot\left[\DFT{M}\V{M}{K}{{\ma{y}_{\text{eq}}}}\right]^T\right)\DFT{M}^H$ & $\hat{\ma{D}} = \frac{1}{M }\DFT{K}\left(\ma{W}_{\text{rx}}\odot\left[\frac{1}{K}\DFT{K}^H\V{K}{M}{\tilde{\ma{y}}_{\text{eq}}}\right]\right)\DFT{M}^H$
		\\
		\hline
		\rule{0pt}{14pt}Conv.	& $\small \IndexM{\frac{1}{K}\hat{\ma{D}}^T \DFT{K}^H}{p}{q} = \sum\limits_{m=0}^{M-1}\IndexM{\V{M}{K}{\bar{\ma{\gamma}}}}{\modulo {p-m}{M}}{q}\IndexM{\V{M}{K}{{\ma{y}}_{\text{eq}}}}{m}{q}$& $\IndexM{\hat{\ma{D}}\DFT{M}} {q}{p} = \frac{1}{K} \sum\limits_{k=0}^{K-1}\IndexM{\V{K}{M}{\bar{\dft{\ma{\gamma}}}}}{\modulo {q-k}{K}}{p}\IndexM{\V{K}{M}{\dft{\ma{y}}_{\text{eq}}}}{k}{p}$\\
		\hline 
		\rule{0pt}{14pt}Rx win.	& $\small \V{M}{K}{\bar{\ma{\gamma}}} = \frac{1}{M}\DFT{M}^H\ma{W}_{\text{rx}}^T$& $\V{K}{M}{\bar{\dft{\ma{\gamma}}}} = \DFT{K}\ma{W}_{\text{rx}}$\\
		\hline 		
	\end{tabular}
\end{table*}
In this section, we represent \ac{GFDM} by means of the discrete Zak transform \cite{ZAK}. This reflects the involvement of \ac{TD} and \ac{FD}-shifted modulation pulses. It also clarifies the structure and facilitates the  implementation.  Consider a vector $\ma{a}\in \compl^{QP\times 1}$, the polyphase matrix of size $Q\times P$ is defined as
\begin{figure}[h]
	\centering
	\includegraphics[width=.8\linewidth]{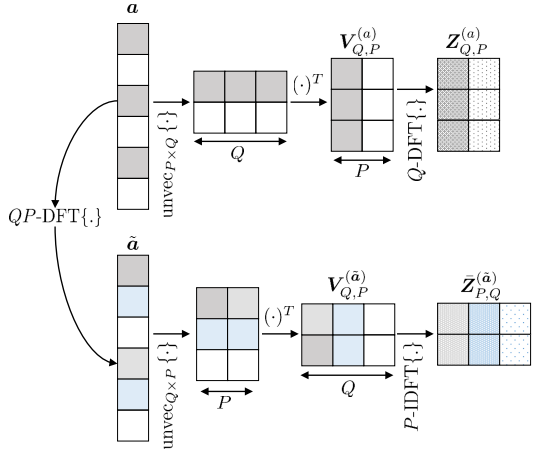}
	\caption{Zak transform. }\label{fig:ZAK_transform}
\end{figure}
\begin{equation}
\small
\V{Q}{P}{\mav{a}} = \unvec{\mav{a}}{P}{Q} ^T\Leftrightarrow\IndexM{\V{Q}{P}{\mav{a}}}{q}{p} = \IndexV{\ma{a}}{p+qP}. \label{eq:polyphase time}
\end{equation}
The $p$-th column of this matrix results from the sampling of $a[n]$ by factor $P$ with shift $p$, as depicted in Fig.~\ref{fig:ZAK_transform}. By applying $Q$-DFT on each column we get the discrete Zak transform
\begin{equation}
\small
\Z{Q}{P}{\mav{a}}= \DFT{Q}\V{Q}{P}{\mav{a}} \in \compl^{Q\times P}.\label{eq:zak time}
\end{equation}
A dual Zak transform in the frequency domain is obtained for the frequency domain vector $\dft{\ma{a}} = \DFT{PQ}\ma{a}$ as 
\begin{equation}
\small
\Zbar{P}{Q}{\dft{\ma{a}}}= \frac{1}{P}\DFT{P}^H\V{P}{Q}{\dft{\ma{a}}} \in \compl^{P\times Q}\label{eq:zak frequency}.
\end{equation}
The basic \ac{TD}-\ac{GFDM} equation can be reformulated in polyphase form by using two indexes $q = 0,\cdots, K-1$ and $p = 0\cdots, M-1$, such that $n =  + pK$. Thereby, 
\begin{align*}
\small
\IndexV{\ma{x}}{q+pK}
&= \sum\limits_{m=0}^{M-1}\sum\limits_{k=0}^{K-1} d_{k,m} g[<q+pK -mK>_N] e^{j2\pi\frac{k}{K} q}.
\end{align*}
Using the polyphase representation given in \eqref{eq:polyphase time}, then 
\begin{equation}
\small
\begin{split}
\IndexM{\V{M}{K}{\ma{x}}}{p}{q}
&=  \sum\limits_{m=0}^{M-1}\IndexM{\V{M}{K}{\ma{g}}}{<p-m>_M}{q}\IndexM{\ma{D}^T\DFT{K}^H}{m}{q}. \label{eq:advanced conv TD}
\end{split} 
\end{equation}
This defines  circular convolution \cite{Circuilant} between the $q$-th column of $\V{M}{K}{\ma{g}}$ and the $q$-th column of $\ma{D}^T\DFT{K}^H$ and it can be expressed in the \ac{FD} with $M$-\ac{DFT} as
\begin{equation*}
\small
\begin{split}
\IndexM{\DFT{M}\V{M}{K}{\ma{x}}}{p}{q} =
\IndexM{\DFT{M}{\ma{V}}_{M,K}^{(\ma{g})}}{q}{q}\cdot \IndexM{\DFT{M}\ma{D}^T\DFT{K}^H}{p}{q},
\end{split}
\end{equation*}
which corresponds to time-domain Zak transform  \eqref{eq:advanced ZAK TD}. Thus,
\begin{equation}
\small
\begin{split}
{\V{M}{K}{\ma{x}}} =
\frac{1}{M K}\DFT{M}^H\left(K\Z{M}{K}{\ma{g}}\odot\left[\DFT{M}{\ma{D}^T\DFT{K}^H}\right]\right). \label{eq:advanced ZAK TD}
\end{split}
\end{equation}
The demodulator performs the inverse steps. Let $\ma{y}_{\text{eq}}$ be the \ac{TD} equalized signal, then 
\begin{equation*}
\small
\hat{\ma{D}} = \frac{1}{M }\DFT{K}\left(\ma{W}_{\text{rx}}\odot \left[\DFT{M}\V{M}{K}{{\ma{y}_{\text{eq}}}}\right]^T\right)\DFT{M}^H.
\end{equation*}
Here $\ma{W}_{\text{rx}} \in \compl^{K\times M}$ is the receive window corresponding to the demodulator prototype pulse. Accordingly, the demodulation convolution in the \ac{TD} is given by
\begin{equation}
\small
\IndexM{\frac{1}{K}\hat{\ma{D}}^T \DFT{K}^H}{p}{q} = \sum\limits_{m=0}^{M-1}\IndexM{\V{M}{K}{\bar{\ma{\gamma}}}}{\modulo {p-m}{M}}{q}\IndexM{\V{M}{K}{{\ma{y}}_{\text{eq}}}}{m}{q}, \label{eq: demodulator convolution TD}
\end{equation}
where $\small \V{M}{K}{\bar{\ma{\gamma}}} = \frac{1}{M}\DFT{M}^H\ma{W}_{\text{rx}}^T$.
A dual \ac{FD} representation can be derived in a similar way. Table \ref{Tab: GFDM basics suammary} summarizes the \ac{GFDM} modem equations for both \ac{TD} and \ac{FD} representations.

%% file: 3_section2.tex
\section{Proposed GFDM modem architecture}\label{sec:prposed archetecture}
\begin{figure*}[t]
	\centering
	\includegraphics[width=1\linewidth]{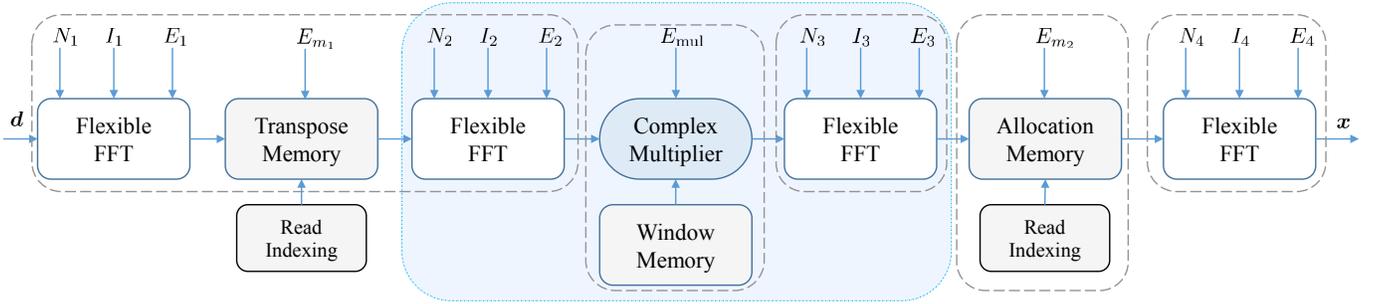}
	\caption{Unified architecture for \ac{TD} and \ac{FD} \ac{GFDM} processing. }\label{fig:Flexible_modulator}
\end{figure*}
Our proposed architecture is developed based on the Zak transform representation of the \ac{GFDM} samples. The \ac{TD} representation can be reformulated as 
\begin{equation*}
\small
\begin{split}
{\V{M}{K}{\ma{x}}} =
\frac{1}{M }\DFT{M}^H\left(\ma{W}_{\text{tx}}^T\odot \left(\DFT{M}\left[\frac{1}{K}\DFT{K}^H\ma{D}\right ]^T\right)\right),
\end{split}
\end{equation*}
and the \ac{FD} representation can be reformulated as 
\begin{equation*}
\small
\begin{split}
{\V{K}{M}{\tilde{\ma{x}}}} =
\DFT{K}\left(\ma{W}_{\text{tx}}\odot \left(\frac{1}{K}\DFT{K}^H\left[\DFT{M}\ma{D}^T\right ]^T\right)\right).
\end{split}
\end{equation*}
Both equations have similar structure with simple differences. In the \ac{TD}, the data symbols are fed to the modulator by the columns of $\ma{D}$.  Then, a $K$-I\ac{DFT} is computed for each column, and the result of $N$ samples need to be stored in a matrix $\dft{\ma{D}}$ of size $K\times M$. Afterwards, the samples of  $\dft{\ma{D}}$ are forwarded row-by-row to an $M$-DFT block. The output of the $M$-\ac{DFT} can be directly element-wise multiplied with the rows of the stored modulation window $\ma{W}_{\text{tx}}$. The result is then fed to an $M$-IDFT block and the output is stored in the columns of the matrix ${\V{M}{K}{{\ma{x}}}}$. Finally, the core block is generated by reading ${\V{M}{K}{{\ma{x}}}}$ in rows. The \ac{FD} modulator, works similarly by feeding the data of $\ma{D}$ in rows, and replacing the $K$ by $M$ and the \ac{DFT} by \ac{IDFT}. The resulting \ac{FD}  samples are stored in a matrix ${\V{K}{M}{\tilde{\ma{x}}}}$ and read by rows.  Additionally, the \ac{FD} modulator requires a final $N$-\ac{DFT} to transform the block to the \ac{TD}. 

The radix-2 values of $K$ and $M$ enables the implementation of \ac{DFT} with flexible \ac{FFT} \ac{IP} cores, e.g. Xilinx \ac{FFT}. This core allows the run-time reconfiguration of the size of \ac{DFT} and setting the block either in \ac{DFT} or  \ac{IDFT} mode.  The proposed  architecture 
is illustrated in Fig.~\ref{fig:Flexible_modulator}. In this design,  $4$ flexible \ac{FFT} cores with the configuration parameter $N_x$ for the \ac{DFT} size and $I_x$ to set the core in the direct $[D]$ or inverse $[I]$ mode. Additionally, each \ac{FFT} block can be enabled $[E]$ or disabled $[D]$ with the parameter $E_x$, such that the disabled block forwards the samples to the next stage. The $4$-th \ac{FFT} core is optional depending on the desired domain of implementation. Moreover, one memory block is used to store the result of the first transform and performs the transpose by the configuration of the indexing unit. The modulator window is stored in a memory as part of the modulator configuration. This memory is always read incrementally and written with respect to the implementation domain. Furthermore, one high throughput complex multiplier is used to perform the element-wise multiplication. A third memory is required to store the samples prior to generating the final block by performing the transpose.  All the  building blocks can be disabled as well. Actually  the highlighted box corresponds to the computation of the convolution. Table \ref{tab: mod Configuration} summarize the configuration parameters of the modulator.
\begin{table}
	\small
	\centering
	\caption{Configuration parameters for modulator.} \label{tab: mod Configuration}
	\begin{tabular}{l|l|l} 		
		Parameter & TD  & FD  \\  
		\hline
		\rule{0pt}{11pt}	$[N_1,N_2,N_3,N_4]$	& $[K,M,M,-]$ & $[M,K,K,N]$  \\ 
		\hline 	
		\rule{0pt}{11pt}	$[I_1,I_2,I_3,I_4]$& $[I,D,I,-]$ & $[D,I,D,I]$ \\ 
		\hline 	
		\rule{0pt}{11pt}	$[E_1,E_2,E_3,E_4]$& $[E,E,E,D]$ &$[E,E,E,E]$  \\ 
		\hline 	
		\rule{0pt}{11pt}	Window & $\ma{W}_{\text{tx}}^T$ &$\ma{W}_{\text{tx}}$  \\ 	
		\hline 	
		\rule{0pt}{11pt}	allocation & $\{\cdot\}^T$ &$\{\cdot\}^T$  \\ 
		\hline 	
	\end{tabular}  
\end{table}
\begin{table}
	\small
	\centering
	\caption{Demodulator configuration, \ac{FD} equalization.} \label{tab: demod Configuration}
	\begin{tabular}{l|l|l} 
		Parameter & TD  & FD  \\ 
		\hline 	
		\rule{0pt}{11pt}	deallocation & $\unvec{\cdot}{K}{M}^T$ &$\unvec{\cdot}{M}{K}^T$ \\ 
		\hline 
		\rule{0pt}{11pt}	Window & $\ma{W}_{\text{rx}}^T$ &$\ma{W}_{\text{rx}}$  \\ 				
		\hline 
		\rule{0pt}{11pt}	$[N_4,N_3,N_2,N_1]$	& $[N,M,M,K]$ & $[-,K,K,M]$  \\ 
		\hline 	
		\rule{0pt}{11pt}	$[I_4,I_3,I_2,I_1]$& $[I,D,I,D]$ & $[-,I,D,I]$ \\ 
		\hline 	
		\rule{0pt}{11pt}	$[E_4,E_3,E_2,E_1]$& $[E,E,E,E]$ &$[D,E,E,E]$  \\ 
		\hline 			
	\end{tabular}  
\end{table}
The demodulator has similar architecture in reversed order. The $N_4$-\ac{FFT} block is enabled or disabled depending on the domain of the equalized signal and the configuration of the demodulator. For instance, if the input is an \ac{FD} equalized signal and the modulator is configured in \ac{TD}, this block needs to be configured as $N$-I\ac{FFT}. The allocation  memory is used to store the received signal in columns and forward it in rows to the next block.  The samples at the output of the demodulator correspond to the columns of $\hat{\ma{D}}$ in the \ac{TD} and to the rows in the \ac{FD} configuration. The demodulator settings with respect to \ac{FD} equalization are listed in Table \ref{tab: demod Configuration}.

Usually, \ac{FD} equalization is used. Thus, the modem can be configured as \ac{TD} modulator and \ac{FD} demodulator. In this way, the overall modulation and equalization processing requires only one $N$-\ac{FFT} for the equalizer. On the other hand, including $N_4$-\ac{FFT}  preserves  the symmetry of the architecture. Therefore, with fast reconfiguration, the modulator can be reconfigured as demodulator. This solution is typical for low-cost time division duplex transmission. The bypass function of each block allows faster processing of certain waveforms. For instance, \ac{OFDM} requires only one I\ac{FFT} to generate \ac{TD} samples. \ac{DFT}-spread-\ac{OFDM} requires only a single \ac{FFT} in the \ac{FD}. Additionally, one or more \ac{FFT} cores can be disabled for certain precoding methods like in \cite{matthe2016precoded}. Furthermore, $N_4$-\ac{FFT} core can be configured with $N_4>N$ to produce over-sampled signal. The allocation  indexing can be altered to implement sort of \ac{FDMA} and multiplex pilot samples. For example, the recent proposed \ac{OTFS} \cite{OTFS_wcnchadani2017orthogonal} is generated by disabling the final transpose.

%% file: 4_section3.tex
\section{Unified architecture for the state of the art implementations} \label{sec:GFDM modem implementation}
\begin{table*}
	\small
	\centering
	\caption{Summary of the settings of the circular convolution architecture. } \label{Tab:state of the art}
	\begin{tabular}{l|l|l} 	
			
        & Time-domain  & Frequency-domain  \\ 
		\hline
		\multicolumn{3}{c}{\rule{0pt}{10pt} Modulation }\\
		\hline 
		\hline
		\rule{0pt}{10pt} Transform 	& $K$-IFFT & $M$-FFT  \\ 
		\hline 	
		\rule{0pt}{14pt} Pulse shape  & $ \IndexM{\bar{\ma{G}}^{(m)}}{:}{p} = K\IndexM{{\V{M}{K}{\ma{g}}}^T}{:}{<p-m>_M}$ & $\bar{\ma{G}}^{(l)} = \ma{1}_{1\times K}\otimes \IndexM{{\V{K}{M}{\tilde{\ma{g}}}}^T}{:}{l}$ \\ 
		\hline 	
		\rule{0pt}{14pt}	Data & $ \bar{\ma{D}}^{(m)} = \ma{1}_{1\times M}\otimes \IndexM{\frac{1}{K}\DFT{K}^H\ma{D}}{:}{m}$ &$\IndexM{\bar{\ma{D}}^{(l)}}{:}{q} = \IndexM{{\DFT{M}\ma{D}^T}}{:}{<q-l>_K}$  \\ 
		\hline 	
		\rule{0pt}{10pt} $N$-IDFT& Disabled & Enabled  \\ 
		\hline 	
		\multicolumn{3}{c}{\rule{0pt}{12pt} Demodulation / \ac{FD} equalization}\\	
		\hline 	
		\hline
		\rule{0pt}{14pt} Pulse shape  & $ \IndexM{\bar{\ma{\Gamma}}^{(m)}}{:}{p} = \IndexM{{\V{M}{K}{\bar{\ma{\gamma}}}}^T}{:}{<p-m>_M}$ & $\bar{\ma{\Gamma}}^{(l)} =\frac{1}{K} \ma{1}_{1\times K}\otimes \IndexM{{\V{K}{M}{\bar{\tilde{\ma{\gamma}}}}}^T}{:}{l}$ \\ 
		\hline 	
		\rule{0pt}{14pt}	Data &  $\small \bar{\ma{Y}}^{(m)} = \ma{1}_{1\times M}\otimes \IndexM{{\V{M}{K}{{\ma{y}}_{\text{eq}}}}^T}{:}{m}$ &$\IndexM{\bar{\ma{Y}}^{(l)}}{:}{q} = \IndexM{{\V{K}{M}{\tilde{\ma{y}}_{\text{eq}}}}^T}{:}{<q-l>_K}$   \\
		\hline 
		\rule{0pt}{10pt} Transform 	& $K$-FFT & $M$-IFFT  \\  
		\hline 	
		\rule{0pt}{10pt} $N$-IDFT& Enabled & Disabled  \\ 
		\hline 
	\end{tabular}  
\end{table*}
\begin{figure*}[h]
	\centering
	\includegraphics[width=1\linewidth]{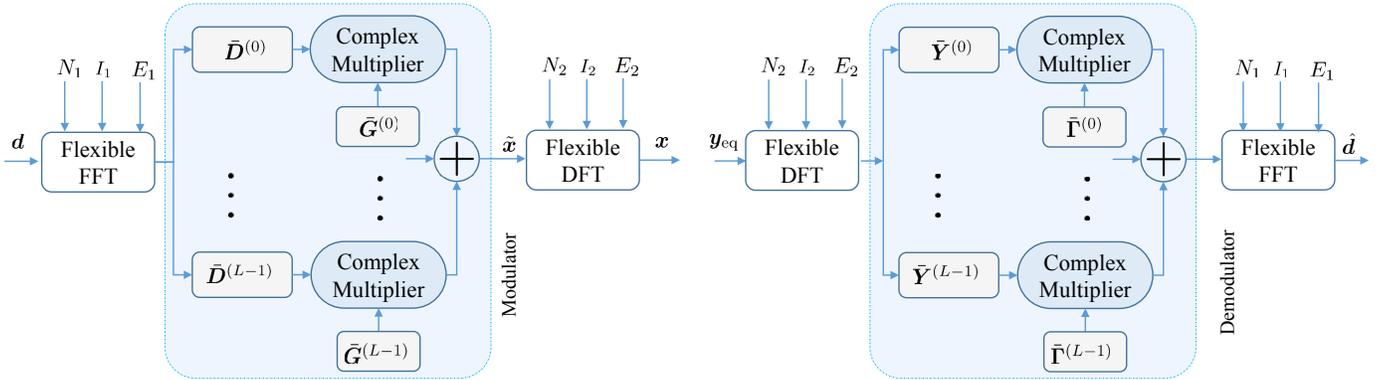}
	\caption{State of the art  implementation.}\label{fig:implementation_state_of_the_art}
\end{figure*}
The state of the art  methods originally target the implementation of the conventional \ac{GFDM} waveform, where mostly periodic \ac{RC} prototype pulse with two-subcarrier overlap  is used. Accordingly, the structure is  influenced by the fact that the parameters $K$  and $M$ cannot be even numbers at the same time for real-valued symmetric pulse \cite{Gabor}. Therefore, to avoid the singularity of the modulation matrix, either $M$ or $K$ need to be odd number. Recently, in \cite{nimr2017optimal}, a detailed study on the prototype filter provides design rules for selecting the proper pulse, and shows that the periodic pulse can be generated by the sampling of the frequency response of a basic filter. The starting point of the samples influences the condition number of the modulation matrix for certain parameters. For even numbers of $K$ and $M$, the starting point has to be shifted by half the sample. This results in a Hermitian symmetric prototype pulse. Moreover, the performance considering this design rule follows the same trend. For instance, the condition number and the noise enhancement increases with the increase of $M$ for a roll-off factor $\alpha >0$. Based on  this restriction, two main methods have been proposed, as described in Section \ref{subsec:TD modem} and Section \ref{subsec: FD modem}.
\subsection{TD Implementation}\label{subsec:TD modem}
The first method focuses on \ac{TD} implementation of \ac{GFDM} modem \cite{danneberg2015flexible}, where $K$ is radix-2 and $M$ is odd. A closer analysis of this architecture shows that this method actually implements the \ac{TD} convolution
\begin{equation*}
\small
\begin{split}
\IndexM{\V{M}{K}{\ma{x}}}{p}{q}
&=\sum_{m=0}^{M-1}\IndexM{\V{M}{K}{\ma{g}}}{<p-m>_M}{q}\IndexM{\ma{D}^T\DFT{K}^H}{m}{q}.
\end{split} 
\end{equation*}
In the first step, the $K$-DFT of the columns of $\ma{D}$ is computed using an I\ac{FFT} core. In order to exploit parallel processing to reduce latency, the convolution is reformulated in the form 
\begin{equation}
\small
\{\V{M}{K}{\ma{x}}\}^T = \sum_{m=0}^{M-1}\bar{\ma{G}}^{(m)}\odot \bar{\ma{D}}^{(m)},
\end{equation}
where $\bar{\ma{D}}^{(m)}$ is used to store $M$ replicas of the \ac{DFT} of the $m$-th column of $\ma{D}$, and $\{\bar{\ma{G}}^{(m)}\}$ are prestored  matrices derived from the prototype pulse. The mathematical details are summarized in Table \ref{Tab:state of the art}. After the computation of the \ac{DFT} of the $M$ columns of $\ma{D}$ and storing the results in $\{\bar{\ma{D}}^{(m)}\}$, $M$ complex multipliers  work in parallel to element-wise multiply the columns of $\{\bar{\ma{D}}^{(m)}\}$ and $\{\bar{\ma{G}}^{(m)}\}$. The outputs of the multipliers  are summed together producing the block samples. The demodulation is performed in a reverse order. The first step is computing the convolution 
\begin{equation*}
\small
\IndexM{\tfrac{1}{K}\hat{\ma{D}}^T \DFT{K}^H}{p}{q} = \sum\limits_{m=0}^{M-1}\IndexM{\V{M}{K}{\bar{\ma{\gamma}}}}{\modulo {p-m}{M}}{q}\IndexM{\V{M}{K}{{\ma{y}}_{\text{eq}}}}{m}{q}, 
\end{equation*}
which is reformulated as 
\begin{equation}
\small
\frac{1}{K}\DFT{K}^H\hat{\ma{D}} = \sum_{m=0}^{M-1}\bar{\ma{\Gamma}}^{(m)}\odot \bar{\ma{Y}}^{(m)}.
\end{equation}
The results after the $K$-\ac{FFT} transform are  the columns of $\hat{\ma{D}}$.
\subsection{FD Implementation}\label{subsec: FD modem}
The methods proposed in \cite{FD-GFDM}  and \cite{FDE} provide an \ac{FD} implementation under a constraint on the \ac{FD} prototype pulse, and $M$ being a radix-2. Namely, $\dft{g}$ spans only $L\ll K$, typically $1$ or $2$, subcarrier spacing. Exploiting these assumptions,  the first step of the implementation is to perform $M$-\ac{DFT} on the rows of $\ma{D}$ by means of \ac{FFT} core. The \ac{FD} convolution  
\begin{equation*}
\small
\begin{split}
\IndexM{\V{K}{M}{\tilde{\ma{x}}}}{q}{p}
=  \sum\limits_{k=0}^{K-1}\IndexM{\V{K}{M}{\tilde{\ma{g}}}}{<q-k>_K}{p} \IndexM{\ma{D}\DFT{M}}{k}{p}
\end{split}
\end{equation*}
is computed with respect to the location of non-zero samples of $\tilde{\ma{g}}$. Accordingly, the convolution is reformulated such that
\begin{equation}
\small
\begin{split}
\{\V{K}{M}{\tilde{\ma{x}}}\}^T
&= \sum\limits_{l\in \set{L}}{ \bar{\ma{G}}}^{(l)}\odot \bar{\ma{D}}^{(l)}.
\end{split}
\end{equation}
Here, $\bar{\ma{G}}^{(l)}$ contains $K$ replicas of the $M$ samples corresponding to the $l$-th non-zero partition of $\dft{\ma{g}}$, and $\set{L}$ denotes the set of non-zero partitions. More details are given in Table \ref{Tab:state of the art}. The matrices $\{\bar{\ma{G}}^{(l)}\}$ are prestored as part of the configuration of the modulator. The output of the $M$-\ac{DFT} is stored in the matrices $\{\bar{\ma{D}}^{(l)}\}$ in $l$-shifted order. After  $M$-\ac{DFT} is completed, the data from $\{\bar{\ma{D}}^{(l)}\}$ and $\bar{\ma{G}}^{(l)}$ are fed to $L$ parallel multipliers and the output thereafter represents the \ac{FD} samples. Finally, an $N$-\ac{IDFT} transform is applied to generate the \ac{TD} samples. The demodulator works in reverse order. Considering \ac{FD} equalizer, first the reformulated  convolution 
\begin{equation}
\small
\DFT{M}\hat{\ma{D}}^T =\sum\limits_{l\in \set{L}}\bar{\ma{\Gamma}}^{(l)}\odot \bar{\ma{Y}}^{(l)}
\end{equation}
is computed, followed by $M$-I\ac{FFT} which produces the estimated rows of $\hat{\ma{D}}$.

Essentially, both methods can be realized with the same hardware architecture as depicted in Fig.~\ref{fig:implementation_state_of_the_art}. One flexible \ac{FFT} is used to perform the first transform, which is configured as $K$-I\ac{FFT} ($M$-\ac{FFT}) for \ac{TD} (\ac{FD}). The convolution is implemented by means of $L$ parallel complex multipliers, $L$ memory blocks to store the $N$ samples of the reformulated pulses and another $L$ memory blocks to store the result of the transform. The transformed data are written in the memories in different ways, Table \ref{Tab:state of the art}. However, reading the data memory can be performed sequentially. A flexible \ac{DFT} core is configured in the inverse mode  in the \ac{FD} modulator and  $N$-\ac{IDFT} block is used 
in the \ac{TD} after the \ac{FD} equalizer. As a consequence of the constraint on even number $M$ and $K$, the \ac{DFT} core needs to be selected based on other odd radices. For instance, Xilinx \ac{DFT} \ac{IP} core consists of  radix-2, radix-3, and radix-5  subcores, and it allows certain predefined configurations that involve at least one odd radix. It is also limited by a maximum length of $N = 1536$.  Accordingly, this architecture can only be configured as \ac{TD} or \ac{FD} modem, which requires in total $2$ \ac{DFT} cores, one for the modem and the other for the \ac{FD} equalization. The configuration parameters are also limited by $L$, the number of parallel multipliers. For instance, the \ac{TD} supports up to $M = L$ and the \ac{FD} modem is only able to work with prototype pulses that do not exceed $L$ subcarrier overlapping. Nevertheless, by replacing the \ac{DFT} core by an \ac{FFT} core, this architecture can be also used for radix-2 parameters.

%% file: 5_section4.tex
\section{Complexity and flexibility analysis}\label{sec:complexity anlysis}
\begin{figure}[h]
	\centering
	\includegraphics[width=.9\linewidth]{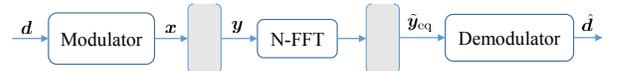}
	\caption{Complexity evaluation model.}\label{fig:complexity_setup}
\end{figure}

In this section, we evaluate  the complexity of the overall modem  considering radix-2 parameters and \ac{FD} channel equalization, where the input signal to the demodulator is in the \ac{FD}, as illustrated in Fig.~\ref{fig:complexity_setup}. The complexity analysis involves the modulator, demodulator, and the $N$-\ac{FFT} transform at the equalizer.
The proposed architecture in Fig.~\ref{fig:Flexible_modulator} denoted here as \emph{\ac{FFT}-based} and the state of the art architecture is called \emph{direct}. This naming is chosen  with respect to the way the circular convolution is processed. The \ac{DFT} core  in Fig.~\ref{fig:implementation_state_of_the_art} is replaced by \ac{FFT}, to enable radix-2 processing. Depending on the configuration of the modulator-demodulator, we get three possible realizations. Namely, \ac{TD}-\ac{FD}, \ac{TD}-\ac{TD}, and \ac{FD}-\ac{FD}. 
\subsection{Number of multiplications}
The $N$-\ac{FFT} requires $\frac{N}{2}\log_2(N)$ \acp{CM}.\footnote{For $N$-\ac{FFT}, $\frac{N}{2}\log_2(N)$ multiplications are required for $N>2$, and for $N = 2$, the \ac{FFT} is performed with addition only.}
The \ac{FFT}-based \ac{TD}-\ac{FD} architecture performs $3M$ times  $K$-\ac{FFT}/I\ac{FFT} and $3K$ times $M$-\ac{FFT}/I\ac{FFT} operations.  Therefore, the overall transforms require $3\frac{MK}{2}\log_2(K) + 3\frac{KM}{2}\log_2(M) + \frac{N}{2}\log_2(N) = 2N\log_2(N)$ \acp{CM}.  This is twice the number of \acp{CM} required by \ac{OFDM} of length $N$. Moreover, $2N$ multiplications are needed for the product with the modulation/demodulation window. 
The direct modulation is achieved with $M$ times $K$-I\ac{FFT} transforms using  \ac{TD} modulator,  $K$ times $M$-\ac{IDFT} with \ac{FD} demodulator. Thus, $\frac{MK}{2}\log_2(K) + \frac{KM}{2}\log_2(M) + \frac{N}{2}\log_2(N) = N\log_2(N)$ \acp{CM} for the transforms. The \ac{TD} direct convolution requires $MN$ and the \ac{FD} requires $KN$ \acp{CM}.
If only \ac{TD} or \ac{FD} modem is applied, two $N$-\ac{FFT}/I\ac{FFT} transforms with $N\log_2(N)$ \acp{CM} are required to transform the signal into the corresponding domain. Thus, for \ac{FFT}-based \ac{TD}-\ac{TD} modem, we additionally need $2[\frac{MK}{2}\log_2(K) + 2\frac{KM}{2}\log_2(M) + N]$, and  the \ac{FFT}-based \ac{FD}-\ac{FD} modem requires additional $2[2\frac{KM}{2}\log_2(M) + 2\frac{MM}{2}\log_2(K) + N]$  \acp{CM}. On the other hand, the direct \ac{TD}-\ac{TD} and \ac{FD}-\ac{FD}  require $2[\frac{MK}{2}\log_2(K) + MN]$, and 
 $2[\frac{KM}{2}\log_2(M) + KN]$ \acp{CM}, respectively. The special case of the direct \ac{FD}-\ac{FD}, where the complexity can be reduced by the consideration of the sparsity of the \ac{FD} prototype pulse, reduces the number of \acp{CM} for the convolution to $2LN$. This special case is useful for the processing of conventional \ac{GFDM} waveform. However, the receiver pulse may overlap with more than $L$ subcarriers, e.g. when zero-forcing demodulator is applied with non-orthogonal modulation matrix. 
\begin{table}
	\small
	\centering
	\caption{Total number of \acp{CM} considering \ac{FD} equalization.} \label{tab:number of multiplications}
	\begin{tabular}{l|l} 
		Implementation method & Total number of  \acp{CM}  \\ 
		\hline 	
		\rule{0pt}{12pt}\ac{FFT}-based, \ac{TD}-\ac{FD} &  $2N\log_2(N) + 2N$ \\ 				
		\hline 		
	\rule{0pt}{12pt}\ac{FFT}-based, \ac{TD}-\ac{TD} &  $2N\log_2(N) + N\log_2(M)+ 2N$ \\ 		
	\hline 		
	\rule{0pt}{12pt}\ac{FFT}-based, \ac{FD}-\ac{FD} &  $2N\log_2(N) + N\log_2(K)+ 2N$ \\ 		
	\hline 	
\rule{0pt}{12pt}Direct, \ac{TD}-\ac{FD} &  $N\log_2(N) + [K+M]N$ \\ 		
	\hline 			
		\rule{0pt}{12pt}Direct, \ac{TD}-\ac{TD} &  $N\log_2(N) + N\log_2(K)+ 2MN$ \\ 		
	\hline 		
	\rule{0pt}{12pt}Direct, \ac{FD}-\ac{FD} &   $N\log_2(N) + N\log_2(M)+ 2KN$ \\ 		
	\hline 			
		\rule{0pt}{12pt}Direct ($L$ overlap), \ac{FD}-\ac{FD} &   $N\log_2(N) + N\log_2(M)+ 2LN$ \\ 		
	\hline 	
	\end{tabular}  
\end{table}
\begin{figure}[h]
	\centering
	\includegraphics[width=1\linewidth]{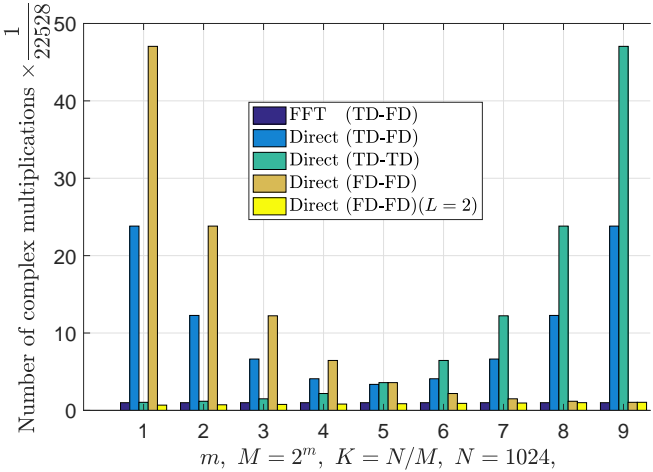}
	\caption{Number of \acp{CM} for different parameters. }\label{fig:Complexity}
\end{figure}
\begin{figure*}[t]
	\centering
	\includegraphics[width=.9\linewidth]{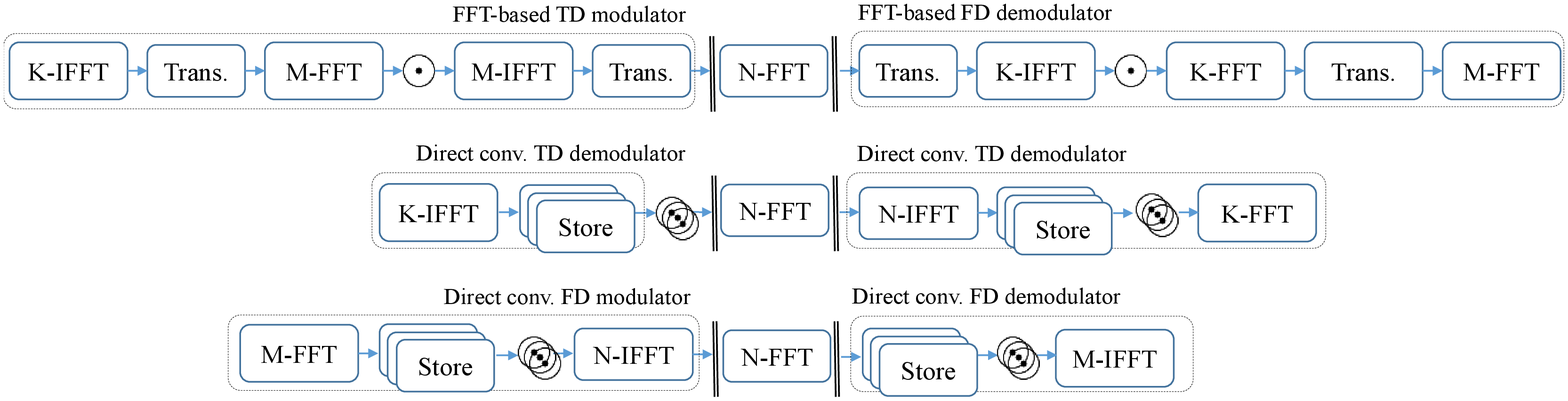}
	\caption{Efficient modem realizations. }\label{fig:Modem_setup}
\end{figure*}

Table \ref{tab:number of multiplications} lists the overall number of \acp{CM} for the different type of modem realizations.  The number of \acp{CM} required by the  \ac{FFT}-based \ac{TD}-\ac{FD} design depends only on $N=KM$ but not on the individual $K$ and $M$. The complexity is  approximately $2$ times the complexity of \ac{OFDM}.  Obviously, the \ac{FFT} based implementation should only consider the combination of \ac{TD} modulation and \ac{FD} demodulation to benefit from the complexity reduction. On the contrary, the complexity of the direct \ac{TD}-\ac{FD} implementation depends on the sum $K+M$, which in most cases requires more \acp{CM} than the one based on \ac{FD} or \ac{TD} only, as can be seen in Fig.~\ref{fig:Complexity}. For $M<K$, it is more efficient to use the direct \ac{TD}-\ac{TD} modem, while in the case  of $K<M$, using the direct \ac{FD}-\ac{FD} modem is more efficient. Thus, the direct convolutional structure can be switched depending on the smaller parameter. Compared  with the \ac{FFT}-based \ac{TD}-\ac{FD}, the complexity of the direct \ac{TD}-\ac{TD} is slightly higher with less than $2$-times up to $M < 16$,  $N=1024$. For $M=16$, it is $2.1$ times higher and For $K=M = 32$, the complexity is $3.6$ times higher. For $M>32$, the modem can be switched to the \ac{FD}-\ac{FD} mode.  In the special direct \ac{FD}-\ac{FD} with $L=2$, the complexity increases with $M$. For $M = N/4$, this modem has the same complexity as the \ac{FFT}-based \ac{TD}-\ac{FD}, and it is about $23\%$ lower for  $M=8$ and $N=1024$. 
\subsection{Hardware analysis }
Based on the complexity analysis, we consider the most efficient setups. Namely,  \ac{FFT}-based \ac{TD}-\ac{FD}, the direct \ac{TD}-\ac{TD} and the direct \ac{FD}-\ac{FD}, as shown in Fig.~\ref{fig:Modem_setup}. The architectures are compared in terms of flexibility, the required resources,  and the modulation/demodulation latency, which also includes the \ac{FD} equalization transform.
Actually, both direct architectures are equivalent, but they differ in the placement of the $N$-I\ac{FFT} block. Let $N_{\text{max}}$ be the maximum supported \ac{FFT} length and $L_{\text{max}}$ the number of parallel multiplication chains in the direct modem.
\subsubsection{Flexibility}
The \ac{FFT}-based architecture supports all combinations of radix-2 $K$ and $M$ with $KM\leq N_{\text{max}}$. The direct architecture supports all combinations that additionally satisfy $\min(K,M)\leq L_{\text{max}}$. For instance, if $N_{\text{max}} = 2048$ and $L_{\text{max}} = 16$, which are the typical values used in \cite{gfdm-flex-modem}, then $(K,M) \in \{ (32,32), (32,64), (64,32)\}$ cannot be supported. Here, it is assumed that the switching between direct \ac{TD}-\ac{TD} and  direct \ac{FD}-\ac{FD} is realized. If the direct architecture is implemented with $N$-\ac{DFT} \ac{IP} cores instead of \ac{FFT} cores, then the allowable combinations are additionally  influenced by the design of the \ac{DFT} core.
On the other hand, the \ac{FFT}-based design can provide additional flexibility by customizing the indexing of the two memories, disabling some blocks, while in the direct convolution this requires more. 
\subsubsection{Resource consumption}
The resource consumption for real-time FPGA implementation is considered in terms of number of consumed \ac{FFT} and complex multiplier \ac{IP} cores, and the \ac{RW} and \ac{R/W} \ac{RAM} blocks. Each block \ac{RAM} can fit up to $N_{\text{max}}$ complex samples. The \ac{RW}-\acp{RAM} are constructed to enable read and write at the same time  for the purpose of pipelining.
Table \ref{tab:resources} lists the number of required resources. 
\begin{table}[h]
	\small
	\centering
	\caption{Number of required resources.} \label{tab:resources}
	\begin{tabular}{c|c|c|c|c} 
		Implementation  & \ac{FFT} & Mult. & (RW)-RAM & R/W-RAM  \\ 
		\hline 	
		\rule{0pt}{12pt}\ac{FFT}-based & $7$& $2$ & $4$ & $2$ \\ 				
		\hline 		
		Direct & $4$& $2L_{\text{max}}$ & $2L_{\text{max}}$ & $2L_{\text{max}}$  \\ 		
		\hline 			
	\end{tabular}  
\end{table}
The \ac{FFT}-based design mainly saves memory and multiplier resources. Only $2$ complex multipliers, $2$ \ac{R/W}-\ac{RAM} blocks are required to store the modulation/demodulation windows, and $4$  \ac{RW}-\ac{RAM} blocks to perform the transpose. The direct architecture  consumes more complex multipliers and memory blocks depending on $L_{\text{max}}$. The \ac{FFT}-based modem requires $3$ more \ac{FFT} \ac{IP} cores. The overall resource consumption depends on the design and implementation of the \ac{IP} cores and the parallel multiplications in the direct architecture. On the other hand, the direct modem requires more resources for control and routing, which is not considered in this evaluation.
\subsubsection{Latency}
In this evaluation, we consider the latency of the modulation/demodulation and the \ac{FD} transform at the equalizer. The latency is a measure of the delay in number of cycles between the first input data symbol at the modulator and the last output symbol at the demodulator. Considering both designs incorporate proper pipelining. Each $N$-\ac{FFT}/I\ac{FFT} requires $N$ cycles to load the samples and $P_M$ cycles to perform the transform. The memory storage of $N$ samples requires $N$ cycles, and the multiplier delay is denoted as $T_{\text{m}}$. Moreover, reading all the demodulated symbols requires $N$ cycles. From Fig.~\ref{fig:Modem_setup}, it can be seen  that  the \ac{FFT}-based implementation requires $6N$ delay cycles corresponding to $4$ memory transposes, the equalizer  $N$-\ac{FFT} load and the unload of the demodulated samples. Moreover, $3(M+K)$ delay cycles to load the data in $3$ time $M$-\ac{FFT}/I\ac{FFT} and $3$ time $K$-\ac{FFT}I\ac{FFT} blocks. The processing delay is $P_N + 3(P_K+P_M) + 2T_{\text{m}}$ for the \ac{FFT}/I\ac{FFT} transforms and multipliers. The overall latency is given by
\begin{equation}
\small
T_{\text{FFT}} =  6N + 3(K+M) +P_N + 3\left( P_K+P_M \right) + 2T_{\text{m}}.
\end{equation}
The direct modem requires $3N$ delay cycles for intermediate storage and unloading the demodulated samples. $2N+2K$ are required for loading the samples to the \ac{FFT}/I\ac{FFT} blocks. The  processing delay is $2P_N+2P_K + 2T_{m}$ for the transforms and the parallel multipliers. Therefore, the overall delay of the direct \ac{TD}-\ac{TD} modem is
\begin{equation}
\small
T_{\text{Direct-TD}} =  5N + 2K +2P_N + 2P_K +  2T_{\text{m}}.
\end{equation}
Similarly for the \ac{FD}-\ac{FD} modem by replacing $K$ by $M$,  
\begin{equation}
\small
T_{\text{Direct-FD}} =  5N + 2M +2P_N + 2P_M +  2T_{\text{m}}.
\end{equation}
The latency is influenced by the  \ac{FFT} size and the configuration of the \ac{FFT} \ac{IP} cores. The higher the \ac{FFT} size, the higher the latency. It is worth mentioning that there is a trade-off between the consumed resources and latency.
\subsubsection*{Numerical example} 
\begin{figure}[t]
	\centering
	\includegraphics[width=1\linewidth]{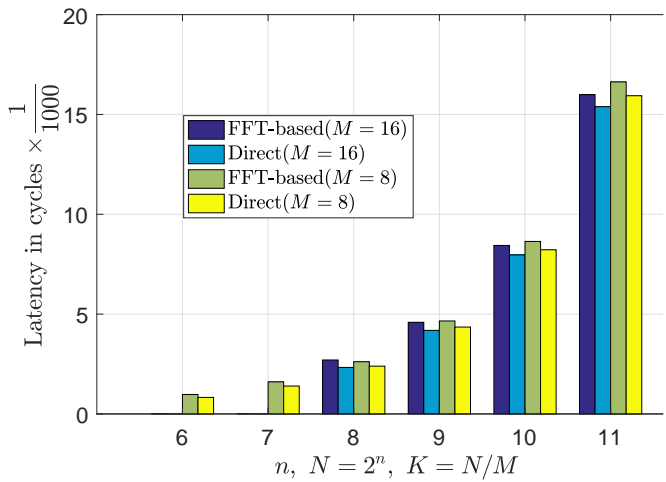}
	\caption{Latency evaluation for different parameters. }\label{fig:Latency}
\end{figure}

\begin{table}[h]
	\small
	\centering
	\caption{Xilinx FFT \ac{IP} processing latency in cycles.}
	 \label{tab:FFT IP prcessing}
	\begin{tabular}{p{10pt}|p{8pt}|p{10pt}|p{12pt}|p{12pt}|p{12pt}|p{12pt}|p{15pt}|p{15pt}|p{14pt}} 
		\hline
		$N$ &$8$ & $16$& $32$ &$64$& $128$ & $256$ & $512$ & $1024$ & $2048$\\ 
		\hline 	
		$P_N$ &$57$ &  $110$& $126$ &$177$& $241$ & $387$ & $643$ & $1170$&$2194$ \\
		\hline 	
	\end{tabular}  
\end{table} 
In this example, the complex multiplier has a latency $T_m = 12$ cycles and the \ac{FFT} blocks are realized based on  an Xilinx-\ac{FFT} \ac{IP} core, which is  configured in the pipelined mode with 3 complex multipliers.  The processing delay of the \ac{FFT} block is listed in Table \ref{tab:FFT IP prcessing}\footnote{In the data sheet of the FFT IP core, the latency includes the number of cycles required to load the input and unload the output, i.e. $2N$ more cycles.}.
The difference between the \ac{FFT}-based and the direct \ac{TD}-\ac{TD} is given by
\begin{equation}
\small
\Delta_{\text{FFT, D-TD}} = N+K+3M+ P_K+3P_M-P_N.
\end{equation}
As shown in Fig.~\ref{fig:Latency} and listed in Table \ref{tab:latency example}, the latency difference  mainly depends on the size of $N$. It can be observed that the \ac{FFT}-based architecture requires additional  latency that decreases with the increase of $N$. For $M=16$, the additional delay decreases from $16.0\%$ at $N = 256$ to $3.9\%$ at $N = 2048$, and for $M=16$, it  decreases from $17.7\%$ at $N = 64$ to $4.3\%$ at $N = 2048$. 
Therefore, the direct design is more appropriate for low latency requirement, especially, with smaller block length. However, this gain is at the cost of significantly increased resources consumption. 
 \begin{table}[h]
 	\small
 	\centering
 	\caption{Latency evaluation.} \label{tab:latency example}
 	\begin{tabular}{c|c|c|c|c|c|c} 
		 \hline 	
		 \multicolumn{7}{c}{$M=16$}\\
		 \hline
 		$N$  & - &-&$256$ & $512$ & $1024$ & $2048$\\ 
 		\hline 	
 		$K$  &-&-& $16$ & $32$ & $64 $&$128$ \\ 	
 		\hline 		
 		$T_{\text{Direct-TD}}$  &- &-&  $2330$& $4186$ & $7966 $&$15390$ \\ 		
 		 \hline
 			$\Delta_{\text{FFT, D-TD}}$  &- &-&  $373$& $405$ & $473 $&$601$ \\ 		
 		 \hline
 Increase $\%$  &- &-&  $16.0$& $9.6$ & $5.9 $&$3.9$ \\ 	
 		\hline 		
 			\hline	
 		 \multicolumn{7}{c}{$M=8$}\\
 		\hline
 		$N$  & $64$ &$128$ &$256$ & $512$ & $1024$ & $2048$\\ 
 		\hline 	
 		$K$  &$8$& $16$ & $32$ & $64 $&$128$ &$256$ \\ 
 		 \hline
 		$T_{\text{Direct-TD}}$  &$828$ &$1398$&  $2394$& $4352$ & $8222$&$15938$ \\ 			
 		\hline 	
 		$\Delta_{\text{FFT, D-TD}}$  &$147$ &$208$&  $222$& $305$ & $418 $&$692$ \\ 	
 			\hline 		
 		  Increase $\%$  &$17.7$ &$14.9$&  $9.3$& $7.0$ & $5.0 $&$4.3$ \\ 	 		
 		\hline 		
 	\end{tabular}  
 \end{table} 

%% file: 6_conclusions.tex
\acresetall
\section{Conclusion}\label{sec:conclusions}
In this work, we provide an overview of multicarrier systems showing that \ac{GFDM} is a main building unit for the multicarrier modulations. \ac{GFDM} can be combined with multiple prototype pulses to develop and optimize new multicarrier waveforms to meet different requirements. \ac{GFDM} block is represented in time and frequency domains as a superposition of parallel circular convolutions. As shown in this paper, the state of the art implementations can be represented in a unified structure that targets the realization of the convolution with parallel multiplier-memory chains. On the contrary, our proposed method realizes the convolution by means of  several \ac{FFT} blocks and only one multiplier. As we demonstrate in this work, the \ac{FFT}-based architecture is computationally more efficient, provides more flexibility, significantly reduces the resource consumption, and maintains  the latency for larger block length.
Additional flexibility can be added to the \ac{FFT}-based architecture with very low  overhead. Namely, the customization of the indexing of the allocation memory provides a ready solution for multiuser scenarios and pilot insertion. Moreover, the bypass function allows faster processing of the classical waveforms in addition to other precoded \acs{OFDM} variants. Furthermore, considering  its fast run-time switching between modulation and demodulation, this architecture provides a low-cost solution for time-division duplex networks. On the other hand, the direct convolution based implementation is a convenient approach  for low latency applications with short block length. However,  even though  the direct architectures employ parallelism, the latency reduction is marginal compared to the \ac{FFT}-based architecture for larger \ac{GFDM} block length. In general, the \ac{FFT}-based implementation is more efficient in terms of number of complex multiplications. Nevertheless, the direct frequency-domain modem is still a reasonable choice for conventional \ac{GFDM} waveform.